\documentclass[notitlepage,a4paper,aps,prd,twocolumn,superscriptaddress,nofootinbib,groupedaddress]{revtex4}
\usepackage{graphicx}
\usepackage[colorlinks=true, pdfstartview=FitV, linkcolor=blue, citecolor=red, urlcolor=magenta]{hyperref}

\usepackage{dcolumn}
\usepackage{graphics}
\usepackage{amssymb}
\usepackage{amsmath}

\usepackage{url}
\usepackage{color}
\usepackage[utf8]{inputenc}
\usepackage{lipsum}

\usepackage{ulem}
\usepackage{tikz}
\usetikzlibrary{arrows,decorations.markings}

\newcommand{\rmd}{\textrm{d}}
\newcommand{\e}{\textrm{e}}

\begin{document}
\newcommand{\cbpf}{
\affiliation{Department of Cosmology, Astrophysics and Fundamental Interacions-COSMO, Centro Brasileiro de Pesquisas F\'{\i}sicas-CBPF, rua Dr. Xavier Sigaud 150, 22290-180, Rio de Janeiro, Brazil.}}
\newcommand{\ufes}{\affiliation{PPGCosmo, CCE - Universidade Federal do Esp\'{\i}rito Santo, Vit\'{o}ria-ES, 29075-910, Brazil.}}
\newcommand{\ufrj}{\affiliation{UFRJ -Universidade Federal do Rio de Janeiro, Instituto de Física, Caixa Postal 68528, Rio de Janeiro, Brazil}}

\title{Bouncing completion of eternal inflation}

\author{Alexsandre L. Ferreira Jr}\email{alexsandrej@gmail.com}
\ufes
\ufrj
\author{Nelson Pinto-Neto}\email{nelsonpn@cbpf.br}
\cbpf
\author{Vanessa N. Xavier}\email{vanessaxavier@cbpf.br}
\cbpf

\date{\today}

\begin{abstract}

Using a purely kinematical argument, the Borde-Guth-Vilenkin (BGV) theorem \cite{N3} states that any maximal space-time with average positive expansion is geodesically incomplete, hence past eternal inflation would be necessarily singular. Recently, discussions about the broadness of this theorem have been resurfaced by applying it to new models and/or challenging the space-time maximality hypothesis. In the present work, we use reference frames of non co-moving observers and their kinematical properties in order to inquire into the nature of such possible singular beginnings. Using the spatially flat de Sitter (dS) space-time as a laboratory, this approach allows us to exhaust all possibilities bounded by the BGV theorem in the case of general spatially flat Friedmann-Lema\^{\i}tre-Robertson-Walker (FLRW) geometries. We show that either there exists a scalar or parallelly propagated curvature singularity, or the space-time must be past asymptotically dS (with a definite non-zero limit of the Hubble parameter when the scale factor becomes null, hence excluding certain cyclic models) in order to be extensible. We are able to present this local extension without violating the null energy condition, and we show that this extension must contain a bounce. This is a mathematical result based on purely kinematical arguments and intuition. The possible physical realization of such extensions are also discussed. As a side product, we present a new chart that covers all de Sitter space-time.

\end{abstract}

\maketitle

\section{Introduction}

In spatially flat Friedmann-Lema\^{\i}tre-Robertson-Walker (FLRW) space-times, in which the scale factor vanishes only when cosmic time goes to minus infinity, free-falling co-moving observers are geodesically complete, experiencing an apparently eternal Universe. However, this happens due to a kinematical effect. In such space-times, non co-moving observers tend to be light-like as the scale factor shrinks, inducing an infinite time dilation of the cosmic time with respect to the finite time measure of the non co-moving clock. These non co-moving geodesics are, therefore, incomplete.

This was perceived by Borde, Guth and Vilenkin, who contrived it into the BGV theorem \cite{N3,Borde_1994}, providing a route to verify incompleteness: if the average expansion of the spatial sections is positive, the space-time is geodesically incomplete and, consequently, singular \cite{Ellis1977}. An alternative definition of singularity has been proposed in \cite{Belinski_2017}.

Recently, the theorem gained renewed attention as cosmologies with cyclic properties, such as periodic Hubble functions and/or conformal cycles \cite{IS19,IS21,P10}, were argued to be incomplete \cite{KS22,N4}. Further, counter-examples were found due to the existence of loopholes in the construction of the average expansion \cite{LED23}, afterwards amended in a re-proposal \cite{GHZ24}. Still, past eternal inflation must be modified and, together with any eternal Universe proposal, can only be complete at the cost of breaking the null energy condition \cite{EL24a,EL24b,carballorubio2024}. However, a fundamental assumption in all approaches is that the space-time geometry is inextensible \cite{N2, Wald1984rg, Penrose65, GEROCH1968526, Hawking66-I, Hawking66-II, Hawking67-III, Hawking-Penrose, Ellis1977, Senovilla_2015}. Thus, the question of whether an averaged expanding space-time can be past eternal depends on whether it admits an extension or not, which enters as an hypothesis in the BGV theorem. Along this route, extendibility conditions for singular space-times were pursued in \cite{Clarke-local-extensions, Clarke-local-extensionsII, Clarke-singularities-hyperbolicST}, and for flat FLRW in \cite{Klein_2017,LP17,YQ18,HISC22,Nomura_2021}, through co-moving coordinates, curvature tensors in parallelly propagated tetrads and/or null-like coordinates for the case of quasi-de Sitter \cite{N5}.
The aim of this paper is to propose a $\mathcal{C}^2$-extension of such space-times through the construction of a physically motivated chart given by the non co-moving observer reference frame.  The $\mathcal{C}^2$ requisite for the extension comes from Einstein's equations, which are second order coupled non-linear partial differential equations for the metric. As an intuitive argument note that, together with the infinite time dilation of the co-moving observer clock, there must be an infinite contraction of their spatial rulers, inducing a zero spatial volume in this asymptotic limit. Hence, the spatial volume seen by the non co-moving observers in this asymptotic limit may be finite, allowing its extension. Indeed, this approach allows us to exhaust all possible cases in flat FLRW space-times, showing that only past asymptotic de Sitter (dS) can be extended (excluding the cyclic models presented in Refs.~\cite{IS19,IS21}). There, the non co-moving observer sees his co-moving volume reaching a velocity dependent finite minimum, before growing again: the extension must contain a bounce. This paper is organized as follows: In Section \ref{sec_BGV} we revisit the BGV theorem, restating its assumptions, some loopholes, and the revised proposal. In section \ref{sec_deSitter} we present a local analytic extension to non co-moving observers in the spatially flat dS space-time. In section \ref{sec_extension} we derive necessary conditions for a $\mathcal{C}^2$ metric extension of an arbitrary spatially flat FLRW space. We conclude in section \ref{sec_conclusion}.

\section{The Borde-Guth-Vilenkin Theorem}
\label{sec_BGV}

The starting point of the BGV theorem is the metric of a spatially flat FLRW:
\begin{equation}
    \rmd s^2=-\rmd t^2+a^2(\rmd r^2+r^2\rmd\Omega^2),
    \label{FLRWmetric}
\end{equation}
the only degree of freedom being the scale factor $a\equiv a(t)$, whose dynamics is determined by an energy-momentum tensor sourcing Einstein's equations. Nonetheless, the following theorem and the upcoming results in this work acquire some of its strength and generality in being purely geometrical results, i.e., by making no reference to the matter content generating space-time curvature or any energy condition. Only in the end of the paper we discuss the implications of our results to the energy-momentum distribution of matter fields using General Relativity.

The geodesic equations are:
\begin{equation}
    \left(\frac{\rmd t}{\rmd s}\right)^2-a^2\left(\frac{\rmd \vec{x}}{\rmd s}\right)^2=\mbox{const},\hspace{5mm}
    a^2\frac{\rmd\vec{x}}{\rmd s}=v_0.
    \label{geoeq}
\end{equation}
The former states that the vector tangent to the geodesics, $u^{\mu}$, has constant norm, either $u^2=-1$ or $u^2=0$, for the time-like or null case, respectively. The latter equation introduces an integration constant, $v_0$, which defines a congruence of geodesics. When $v_0=0$ the geodesics are the trajectories of co-moving observers w.r.t. the cosmological background. The fractional expansion of such congruence, as seen by a geodesic ``observer" $u_0^{\mu}$, with affine parameter $s$, is given by the generalized Hubble factor, defined as
\begin{equation}\label{Hbgv_general}
    H_{BGV}(s)=\frac{-u^\mu_0}{\sigma^2+\kappa}\frac{D u_\mu}{\rmd s}=\frac{-1}{\sigma^2+\kappa}\frac{\rmd \sigma}{\rmd s}=\frac{\rmd}{\rmd s}\mathcal{F}(\sigma),
\end{equation}
being $\sigma=u_0^{\nu}u_{\nu}$, $\kappa=u_0^{\nu}u_{0\nu}$, and $D/\rmd s \equiv u_0^{\nu}\nabla_{\nu}$ is the absolute derivative along the observer's worldline (note that the vector with the lower zero index $u_0^{\mu}$ is the observer measuring the expansion of the congruence given by $u^{\mu}$, with no additional indexation). For a time-like $u^\mu_0$, $\sigma=(1-v_{rel}^2)^{-1/2}$, i.e., represents the Lorentz factor for the relative spatial velocity $v_{rel}$ between both observers, and $\kappa=-1$, while for a light-like $u^\mu_0$, $\sigma=\rmd t/\rmd s$, the change of time w.r.t. the affine parameter, and $\kappa=0$. Moreover,
\begin{equation}
    \mathcal{F}(\sigma)=
    \begin{cases}
      \sigma^{-1} & \text{when $\kappa=0$,}\\
      \frac{1}{2}\mathrm{ln}\frac{\sigma+1}{\sigma-1} & \text{when $\kappa=-1$.}\\
    \end{cases}       
\end{equation}

To grasp some physical intuition into the significance of $\mathcal{F}$, note that, given a time-like observer $u^{\mu}_0$ and a non-relativistic relative velocity, $v_{rel}=|\vec{u}|\ll1$, we find that
\begin{equation}
    \mathcal{F}\approx-\mbox{ln}\,v_{rel}.
\end{equation}
Hence,
\begin{equation}
    H_{BGV}\approx-\frac{\rmd}{\rmd s}\mbox{ln}\,v_{rel}=\frac{\Delta u_r}{\Delta r},
\end{equation}
where $\Delta r=|\Delta \vec r|=|\vec{u}\Delta s|$ is the separation of two test particles in the congruence $u^{\mu}$ by elapsed time $\Delta s$, and $\Delta u_r=\Delta\vec{u}\cdot\Delta \vec{r}/\Delta r$ is the radial projection of $\Delta\vec{u} \equiv \vec{v}_{rel} =-(\rmd\vec u/\rmd s)\Delta s$, which is the relative velocity between test particles. As a consequence, $H_{BGV}$ is just the fractional change in the velocity of the congruence $u^{\mu}$, as expected (see Ref.~\cite{N3} for details).

Alongside that, the generalized Hubble parameter reduces to the usual one $H=(1/3)u^{\mu};_\mu$ when smeared over all directions of $u^{\mu}_0$ \cite{N3}. Further, defining the averaged expansion rate $\mathcal{H}_{avg}$ as
\begin{equation}
    \mathcal{H}_{avg}\int^{s}_{s_i}\rmd s=\int^{s_f}_{s_i}H_{BGV}\,\rmd s,
    \label{Havg}
\end{equation}
where $\Delta s=s_f-s_i$ is the parameter interval, the BGV theorem states that, if the averaged expansion rate is positive (negative), the space-time is geodesically past (future) incomplete \cite{N3}.

The proof follows by direct integration of the generalized Hubble factor, giving
\begin{equation}
    \int^{s_f}_{s_i}H_{BGV}\,\rmd s=\mathcal{F}(\sigma_f)-\mathcal{F}(\sigma_i)\leq \mathcal{F}(\sigma_f).
\end{equation}
Then, if $\mathcal{H}_{avg}>0$:
\begin{equation}
    0<\mathcal{H}_{avg}\leq\frac{\mathcal{F}(\sigma_f)}{\Delta s},
\end{equation}
which means that $s_i>-\infty$, and that the space-time is past incomplete. Future incompleteness proceed analogously for $\mathcal{H}_{avg}<0$.

In Ref. \cite{LED23}, counter-examples to the BGV theorem were constructed -- complete space-times where $\mathcal{H}_{avg}>0$ --  due to the fact that the interval $\Delta s$ and, consequently, $\mathcal{H}_{avg}$, are not well defined for complete geodesics, i.e., when $s_i\to-\infty$ or $s_f\to\infty$. Such loophole can be amended by redefining the averaged expansion as
\begin{equation}
    \mathcal{H}_{0}=\frac{1}{s_f-s_0}\int^{s_f}_{s_0}H_{BGV}\,\rmd s,
    \label{Havg}
\end{equation}
where now $s_0\in (s_i,s_f)$, which cannot be taken strictly equal to minus infinity. Thence, whenever exists $\Delta>0$, so that $\mathcal{H}_{0}\geq \Delta$ for all $s_0\in (s_i,s_f)$, the space-time is past incomplete, without the presence of loopholes, see Ref. \cite{GHZ24} for details.

Nonetheless, the present work intends to scrutinize a different aspect of the BGV theorem, namely, the assumption that the space-time is inextensible. Important works have been done in this direction \cite{Klein_2017,LP17,YQ18,HISC22,Nomura_2021}. However, here we seek a different approach by constructing a chart adapted to the incomplete non co-moving observer, allowing us to grasp some physical intuition on possible extensions and the elusive nature of some singular space-times.

\section{de Sitter laboratory case}
\label{sec_deSitter}

 The de Sitter space-time, a maximally symmetric solution to Einstein's equations with positive cosmological constant \cite{N2}, is a pivotal example. The coordinate chart in which the spatial section are flat homogeneous and isotropic hypersurfaces, given by eq. (\ref{FLRWmetric}) with an exponential scale factor $a(t)=a_{dS}(t)=\e^{\alpha t}$, satisfies all the BGV theorem hypothesis and is therefore incomplete, albeit the absence of any singularity, indicating that the incompleteness is just a limitation of the flat patch.

The completeness is manifested in the closed dS foliation with topology $\mathbb{R}\times \mathbb{S}^{3}$, in which the homogeneous isotropic spatial hypersurfaces are closed. Nonetheless, this coordinate change resorts to the global structure of the manifold. Therefore, aiming to extend more general space-times, that are approximately but not exactly dS, we pursue in this section an extension in which the incomplete non co-moving time-like geodesics are the coordinate curves, i.e., a chart in which such observer is at rest, giving an intuitive understanding of its fate.

The new time will be the proper time of the observer and $u^{\mu}=(u^t,u^r,0,0)$, the tangent vector to the geodesics in eq. (\ref{geoeq}), given by

\begin{equation}
    u^t=\frac{\rmd t}{\rmd \tau}\equiv\gamma=\sqrt{1+\frac{v^2_0} {a^2}},\hspace{5mm}u^r=\frac{\rmd r}{\rmd \tau}=\frac{v_0}{a^2}.
    \label{tangent_non-comoving_observer}
\end{equation}
Note that Eq.~\eqref{deltatau} with $t\to -\infty$ yields $\Delta\tau={\rm arcsinh}(e^{\alpha t_0}/|v_0|)/\alpha$, and flat dS is past incomplete.

The vector $u^{\mu}$ defined in \eqref{tangent_non-comoving_observer} describes radial motion. In the $t-r$ planes we can construct the curves orthogonal to the geodesic congruence, parametrized by $l$. A convenient choice implies in the following coordinate transformations:

\begin{align}
     \mbox{e} ^{\alpha t}&=a_{dS}(\tau,l)=|v_0|\mbox{sinh}(\Theta-\Theta_0) ,
     \label{new a}\\
 r&=l-\frac{1}{\alpha v_0}\left[\mbox{coth}(\Theta-\Theta_0)+\mbox{coth}\Theta_0\right],
    \label{lcoord}
\end{align}
with $\Theta=\alpha(\tau+v_0 l)$ and $\Theta_0=-{\rm arcsinh}(e^{\alpha t_0}/|v_0|)$. The boundary $a_{dS}=0$ of the flat patch is when $\Theta=\Theta_0$, occurring at different proper times on each geodesic, as they have different values of $l$. 

Equations (\ref{new a},\ref{lcoord}) give the coordinate transformations from the co-moving coordinates to the non co-moving ones. Direct derivation w.r.t. the parameter $l$ leads to the normalized vector field $v^{\mu}=(v^{t},v^{r},0,0)$ tangent to the space-like hyper-surfaces in the new coordinate system:

\begin{equation}
    v^{t}=\frac{\rmd t}{\rmd l}=v_0\sqrt{1+\frac{v^2_0}{a^2}},\hspace{2mm}v^{r}=\frac{\rmd r}{\rmd l}=1+\frac{v^2_0}{a^2}.
    \label{spatialtrans}
\end{equation}
 
The line element in the new coordinates reads

\begin{equation}
    \rmd s^2=-\rmd\tau^2+(a^2+v^2_0)\rmd l^2+(ar)^2\rmd\Omega^2.
    \label{newlinelement}
\end{equation}
where $a\equiv a_{dS}(\tau,l)$ and $r\equiv r(\tau,l)$ as in Eqs.~(\ref{new a}) and (\ref{lcoord}). Notice that, given a value of $v_0$ and requiring the motion of the geodesics to occur in the $t-r$ plane, the congruence given by $u^\mu$ and, consequently, the proper time $\tau$, is defined and unique up to a time translation. Nonetheless, the coordinates in the surfaces orthogonal to $u^\mu$ are arbitrary and can be changed by different choices of parametrization. 

Further, although $a_{dS}$ is expressed in terms of a hyperbolic sine, it depends on $l$ also, and the metric above is spatially inhomogeneous and anisotropic. Hence, it cannot be understood as the incomplete dS with negative spatial curvature with a zero volume element when $a=0$, as it was done in Ref.~\cite{N4}, which is clearly not the case here. Note that $l$ is not a radial coordinate. 

The line element (\ref{newlinelement}) ascertain that the $a_{dS}=0$ limit is reached with a finite proper time $\alpha\tau=\Theta_0-\alpha v_0 l$. However, as anticipated previously, due to the same kinematical effects the spatial volume element does not vanish. Indeed, near $\Theta=\Theta_0$, equivalent to $t\to-\infty$, the co-moving volume goes as

\begin{equation}
    \sqrt{-g}\approx \frac{|v_0|\mbox{sin}\theta}{\alpha^2}\,\mbox{cosh}^2(\Theta-\Theta_0),
    \label{vol1}
\end{equation}
which reaches a minimum $\sqrt{-g}=|v_0|\mbox{sin}\theta/\alpha^2$ when $\Theta=\Theta_0$, growing again for $\Theta<\Theta_0$. This is expected as the dS manifold can be covered by two flat patches, and if we assume that time for the non co-moving observer goes only in one direction, one sheet is contracting while the other expands. This is in contrast with Ref. \cite{N6}, in which the complete dS space-time is covered by two expanding sheets, and time runs in different directions and no flow of particles between the sheets must be imposed as a boundary condition. The argument is that the energy density of test particles traveling in the non co-moving geodesics, i.e., the $T^{0}_0$ component of its energy-momentum tensor, diverges in the boundary. However, this is only a coordinate problem. For the co-moving observer the non co-moving one approaches light velocity near this boundary, appearing to have infinity energy, but nothing happens from the point of view of non co-moving observers themselves, analogous to what happens near a black hole's horizon, where observers can naturally pass through it from their own point of view.

Note that Eq.~\eqref{vol1} immediately implies that the expansion of the congruence reads

\begin{multline}
    u^{\mu}_{;\mu}=\alpha\mbox{tanh}(\Theta-\Theta_0)+\\2\alpha\left[\frac{(\alpha v_0 l+\mbox{coth}\Theta_0)\mbox{coth}(\Theta-\Theta_0)-1}{\alpha v_0 l-\mbox{coth}(\Theta-\Theta_0)+\mbox{coth}\Theta_0}\right].
\end{multline}
When $\Theta=\Theta_0$ the expansion becomes $\theta=2\alpha(\alpha v_0 l+\mbox{coth}\Theta_0)$, there is no caustic, as expected. The caustic only occurs when $r=0$, as the coordinate curves meet, by definition.

It is straightforward to see that a contracting flat dS patch, given by $a'_{dS}=\mbox{e} ^{-\alpha t'}$, can be mapped into the metric (\ref{newlinelement}) for $\Theta<\Theta_0$. Performing the same coordinate change given by (\ref{tangent_non-comoving_observer}) and (\ref{spatialtrans}), we find

\begin{align}
   \mbox{e} ^{-\alpha t'}&=-|v_0|\mbox{sinh}[\alpha(\tau+v_0l)-\alpha(\tau_0+v_0l_0)+\Theta_0].\\
 r'&=l-l_0-\frac{1}{\alpha v_0}\left[\mbox{coth}(\Theta-\Theta_0)-\mbox{coth}\Theta_0\right].
    \label{lcoord2}
\end{align}
In order to connect these expression with Eqs.~(\ref{new a}) and Eq.~(\ref{lcoord}), the initial conditions must satisfy $\alpha(\tau_0+v_0l_0)=2\Theta_0$ and $l_0=2 \mbox{coth}\Theta_0 /(\alpha v_0) $. Furthermore, the contracting nature of the dS manifold can be seen by evaluating $H_{BGV}$ given by eq. (\ref{Hbgv_general}). Considering the ``observer" appearing in the definition to be the co-moving one, $u_0^{\mu}=(1,0,0,0)$, then

\begin{equation}\label{HBGV_deSitter}
\begin{gathered}
    H_{BGV} = \frac{a^{2}}{v_0^2} \frac{\rmd \gamma}{\rmd t} = \alpha \tanh (\Theta - \Theta_0).
\end{gathered}    
\end{equation}
Thence, for $\Theta < \Theta_0$ the non co-moving congruence is in the contraction phase, reaching null expansion when $\Theta = \Theta_0$. After crossing the $a=0$ boundary, it enters the expanding phase. Additionally, when $\Theta-\Theta_0\gg1$ ($\Theta-\Theta_0\ll-1$) we have that  $H_{BGV}=\alpha$ ($H_{BGV}=-\alpha$), the usual dS expanding (contracting) Hubble parameter, where co-moving and non co-moving observers coincide, as can be seen by transformations (\ref{tangent_non-comoving_observer}).

\section{Extension for a general flat FLRW metric}
\label{sec_extension}

We focus on past-incomplete space-times. Evidently, if the domain of the cosmological time does not go up to minus infinity, the space-time is already incomplete, and a completion is only possible in the absence of scalar curvature singularities. In a FLRW space-time, there are only three independent scalar quantities, namely: 

\begin{equation}
    \begin{gathered}
        R =R^{\mu}_{\,\,\mu} = 6(\dot{H}+2H^2),\\
        K =R^{\mu\nu\lambda\rho}R_{\mu\nu\lambda\rho}= 12[\dot{H}^2 + 2H^2(\dot{H}+H^2)],\\
        R^{\mu\nu}R_{\mu\nu}= 3[(3H^2+2\dot{H})^2+3 H^4],
    \end{gathered}
\end{equation}
the Ricci and Kretschmann scalars, and the Ricci contracted with itself, respectively. With that, our first assumption is:
 
\textbf{Assumption (i)}: we only consider spatially flat FLRW models where cosmological time can be extended to minus infinity, i.e., $t\in (-\infty,t_0]$, with a vanishing scale factor $\lim_{t\rightarrow-\infty}a(t)=0$. For $\mathcal{C}^{2}$-extendibility we also assume that there the curvature scalars, which are composed by the Hubble function, $H=\dot{a}/a$, and its first time derivative, $\dot{H}$, are all bounded.

On the contrary, if $t\in [t_{in},t_0]$, with $t_{in}$ finite, either $a(t_{in})=0$ and there is a singularity, as in any Big Bang model, or $a(t_{in})=a_{in}\neq0$ and it is trivially extensible, while if $\lim_{t\rightarrow-\infty}a(t)\neq0$, the space-time is past complete \cite{N5}.

Incompleteness in such apparent eternal universes appears from kinematical effects, i.e., by investigating non co-moving geodesics given by the equations in (\ref{geoeq}).

\textbf{Definition 1.} \textit{A flat FLRW space-time, in which $t\in (-\infty,t_0]$, is time-like past-incomplete if the proper time}

\begin{equation}
    \Delta\tau(t)=\int_{t}^{t_0}\frac{\rmd t'}{\sqrt{1+\frac{v^2_0}{a^2}}},
\label{deltatau}
\end{equation}
\textit{has a definite limit $\Delta\tau\to\tau_{*}$ as $t\to-\infty$, being $\tau_*\in\mathbb{R}$, and light-like incomplete if the affine parameter}

\begin{equation}
    \Delta\lambda(t)=\int_{t}^{t_0}a\rmd t',
\end{equation}
\textit{has a definite limit $\Delta\lambda\to \lambda_{*}$ as $t\to-\infty$, being $\lambda_*\in\mathbb{R}$.}

In the cases of interest here, time and light-like past-incompleteness are equivalent, and incompleteness imposes an important kinematical consequence, as shown in the following lemmas:

\textbf{Lemma 1.} \textit{Given a flat FLRW space-time that satisfies assumption \textbf{(i)}, it is time-like past-incomplete if and only if it is light-like incomplete.}

\textit{Proof.} Due to assumption \textbf{(i)}, $a\rightarrow0$ for $t\to-\infty$, hence we can choose $t_0$ such that $a^2(t_0)\ll v^2_0$. Therefore

\begin{equation}
    \lim_{t\to-\infty}\Delta\tau\approx\frac{1}{|v_0|}\lim_{t\to-\infty}\Delta\lambda=\frac{1}{|v_0|}\lim_{t\to-\infty}\left[F(t)\right]\bigg|^{t_0}_t,
    \label{pastincomp}
\end{equation}
being $F(t)$ the primitive of $a(t)$. The space-time is incomplete if $\lim_{t\to-\infty}F(t)\rightarrow F^{*}$, being $F^{*}\in\mathbb{R}$. Thus, when it is light-like incomplete, the limit exist and time-like incompleteness follows, the converse being also true.

\textbf{Lemma 2.} \textit{Let a flat FLRW space-time that satisfies assumption \textbf{(i)}. If it is past-incomplete, then  $H/a$ diverges in the past-boundary}.

\textit{Proof:} Assuming the space-time to be incomplete means that $F(t)\rightarrow F^*$ when $t\to-\infty$, from equation (\ref{pastincomp}). Assuming further that $H/a$ has a well-defined limit $\lim_{t\to-\infty}H/a= d$, being $d\in\mathbb{R}$, we run into a contradiction:

\begin{align}
    \lim_{t\to-\infty}F&=\lim_{t\to-\infty}\frac{Fa}{a}=\lim_{t\to-\infty}\frac{a^2+\dot{a}F}{\dot{a}}\nonumber\\&=\lim_{t\to-\infty}\left(F+\frac{a}{H}\right)=F^{*}+\frac{1}{d}\neq F^*,
    \label{lemma1}
\end{align}
where we have used the l'Hôpital rule in the second equality. As no real number satisfies the condition $1/d=0$, the limit $F(t)\rightarrow F^{*}$ for $t\rightarrow-\infty$ can be a real number only if $H/a\rightarrow\pm\infty$ there, and the Lemma is proved.

Motivated by the de Sitter construction, we apply the coordinate transformation induced by Eqs.~(\ref{tangent_non-comoving_observer}) and (\ref{spatialtrans}) in the radial planes $(t,r)\rightarrow(\tau,l)$ to a general case. The transformed line element from Eq.~\eqref{FLRWmetric} is given by Eq.~\eqref{newlinelement}, with generic $a$.

In the new coordinates $t\equiv t(\tau+v_0l)$ and, consequently $a(t)\equiv a(\tau+v_0l)$. We denote $\xi=\tau+v_0l$, and $\xi_*$ the value of the argument such that $t\rightarrow-\infty$ and $a\rightarrow0$ for $\xi\rightarrow\xi_*$. 

The metric determinant is
\begin{equation}\label{new_metric_determinant}
    \sqrt{-g}=\sqrt{a^2+v_0^2}(ar)^2\mbox{sin}\theta.
\end{equation}
Thence, the condition for the metric components, as well as its determinant, to be well-behaved when $a\rightarrow0$, i.e. to be continuous and non-degenerate at these points, reads
\begin{equation}
    \lim_{\xi\to\xi_*} ar = \lim_{\xi\to\xi_*} \left(al+v^2_0a\int^{l}_0\frac{\rmd l'}{a^2}\right),
\end{equation}
which must not diverge neither vanish. Using $\rmd l = \rmd t /(v_0\gamma)$, $t\to-\infty$ and $t_0$ such that $a^2(t_0)\ll v^2_0$  we get

\begin{equation}
    \lim_{\xi\to\xi_*} ar=\frac{v_0}{|v_0|}\lim_{t\to-\infty}a(t)\left[G(t)\right]\Big|^{t}_{t_0},
    \label{lim_ar}
\end{equation}
being $G(t)$ the primitive of $1/a(t)$, which is finite only if $\dot{a}$ diverges, yielding scalar curvature singularities in this limit. Hence, $G$ must diverge.

\textbf{Lemma 3.} \textit{Given a FLRW model in the non-co-moving coordinates (\ref{newlinelement}), if the functions $a(\xi)$ and $\dot{a}(\xi)$ have well defined limits in the past boundary $(\tau+v_0l)=\xi\to\xi_*$, then}

\begin{equation}
    \lim_{\xi\to\xi_*} ar=-\frac{v_0}{|v_0|}\lim_{t\to-\infty}\frac{1}{H}.
\end{equation}

\textit{Proof:} Given that $\dot{a}$ has a definite limit, G diverges, and we get

\begin{equation}
    \lim_{t\to-\infty}Ga=\lim_{t\to-\infty}\frac{G}{1/a}=-\lim_{t\to-\infty}\frac{1}{H},
\end{equation}
Equation (\ref{lim_ar}) then reads

\begin{equation}
    \lim_{\xi\to\xi_*} ar=-\frac{v_0}{|v_0|}\lim_{t\to-\infty}\frac{1}{H}.
\end{equation}

This result implies the second assumption for extensibility:
\vspace{0.15cm}

\textbf{Assumption (ii)}: $H\to\mbox{const}\neq0$ as $t \rightarrow -\infty$.

\vspace{0.15cm}
Furthermore, when assumptions \textbf{(i)} and \textbf{(ii)} hold, then we must have that $\lim_{t\rightarrow-\infty}\dot{H}=0$, see Lemma 2.6 in Ref.~\cite{N5}.

The metric determinant at the past-boundary is given by
\begin{equation}
    \lim_{\xi\to\xi_*}\sqrt{-g}= v_0\mbox{sin}\theta \lim_{t\to-\infty}\frac{1}{H^2}.
\end{equation}

Hence, if assumptions \textbf{(i)} and \textbf{(ii)} are satisfied, the metric (\ref{newlinelement}) gives a $\mathcal{C}^{0}$ extension for past-incomplete space-times. 

If the metric has a dynamical origin from Einstein's equation, it must be at least $\mathcal{C}^2$. For this condition to be met it is necessary and sufficient that the derivatives w.r.t. the new coordinates of the metric elements, up to second order, be finite in the past boundary.

For the spatial $g_{ll}$ element we get

\begin{align}
    \lim_{\xi\rightarrow\xi_*}\partial_\tau(a^2+v_0^2)&=\frac{1}{v_0}\lim_{\xi\rightarrow\xi_*}\partial_l(a^2+v_0^2)=0,\\
   \lim_{\xi\rightarrow\xi_*}\partial^2_\tau(a^2+v_0^2)&=\frac{1}{v^2_0}\lim_{\xi\rightarrow\xi_*}\partial^2_l(a^2+v_0^2)=2v_0^2\left(\dot{H}+H^2\right).
\end{align}

For the angular components we obtain

\begin{multline}
\lim_{\xi\rightarrow\xi_*}\partial_\tau(ar)^2=\lim_{\xi\rightarrow\xi_*}\frac{1}{v_0}\partial_l(ar)^2\\=\lim_{\xi\rightarrow\xi_*}2(ar)\left[\sqrt{a^2+v_0^2}\frac{H}{a}(ar)+\frac{v_0}{a}\right]=-2v_0l
\end{multline}
and

\begin{align}
    \lim_{\xi\rightarrow\xi_*}\partial^2_\tau(ar)^2&=2\lim_{\xi\rightarrow\xi_*}\left[(v_0Hl)^2+\frac{v^2_0}{H^2}\frac{\dot{H}}{a^2}+\frac{\dot{H}}{H^ 2}+1\right], \label{conditionIII}\\
   \lim_{\xi\rightarrow\xi_*}\partial^2_l(ar)^2&=2v^2_0\lim_{\xi\rightarrow\xi_*}\partial^2_\tau(ar)^2-4v_0^2\\
   \lim_{\xi\rightarrow\xi_*}\partial_l\partial_\tau(ar)^2&=2v_0\lim_{\xi\rightarrow\xi_*}\partial^2_\tau(ar)^2-2v_0.
   \label{angular2}
\end{align}

One can see from \eqref{conditionIII} that a new condition emerges: 

\vspace{0.15cm}
\textbf{Assumption (iii)}: $\lim_{\xi\rightarrow\xi_*}\dot{H}/a^2=c$, being $c\in\mathbb{R}$.

\vspace{0.15cm}
If assumption \textbf{(iii)} is not satisfied, the geometry contains a parallelly propagated curvature singularity, see Ref.~\cite{YQ18}, which forbids completion.

Note that the $\mathcal{C}^2$ requirement for the metric $g_{ll}$ component implies that $\partial_\xi (a^2)=0$ and $\partial^2_\xi (a^2)=2v_0^2H^2>0$, for $\xi=\xi_*$, which is a local minimum. Hence, the extension must necessarily contain a bounce. This can be seen through the generalized Hubble factor defined in equation (\ref{Hbgv_general}) between a co-moving observer $u_0^{\mu} =(1,0,0,0)$ and a non co-moving congruence $u^{\mu}=(\gamma,v_0/a^{2},0,0)$:

\begin{equation}
    H_{BGV} = \frac{aH}{\sqrt{a^2 + v_0^2}}.
\end{equation}
Since $\partial_\tau( a^2) =\partial_\xi (a^2) = 2a \partial_{\tau}a$, the expansion can be rewritten as

\begin{equation}
    H_{BGV} = \frac{a H \sqrt{a^2 + v_0^2}}{(a^2 + v_0^2)} = \frac{a \partial_{\tau}a}{(a^2 + v_0^2)} = \frac{1}{2} \frac{\partial_{\xi}(a^2)}{(a^2 + v_0^2)},
\end{equation}
going from negative to positive values, passing through zero in $\xi=\xi^{*}$, characterizing the bounce. What happens before the bounce will depend on the specific model under consideration.

There exists a class of incomplete FLRW space-times that satisfies condition \textbf{(i)}, but $\lim_{t\rightarrow-\infty}H=0$ (like in some pre-Big Bang models \cite{Gasperini_2003, Rinaldi_2005, Gasperini_1994, Gasperini_2000}), being not contemplated by our extension because of the requirement that the volume element be finite, leading to condition \textbf{(ii)}. Such failure could be simply a coordinate problem: our choice does not fit for such cases. However, the divergence of the volume element due to the angular component means that observers in different planes of motion diverge from each other in a finite proper time, which might signal a true problem. Indeed:

\textbf{Lemma 5.} \textit{Let a spatially flat FLRW space-time, which satisfies assumption \textbf{(i)}. If $H\to0$ in the past-boundary, then $H^2/a^2\to\dot{H}/a^2$ as $t\to-\infty$.}

\textit{Proof:} As $H\to0$ we have directly by using the l'Hôpital rule that

\begin{equation}
    \lim_{t\to-\infty}\frac{H^2}{a^2}=\lim_{t\to-\infty}\frac{H\dot{H}}{a\dot{a}}=\lim_{t\to-\infty}\frac{\dot{H}}{a^2}.
\end{equation}

\textbf{Corollary 1.} \textit{Let a spatially flat FLRW space-time satisfying assumption \textbf{(i)}. If $H\to0$ the space-time is either complete, or $\dot{H}/a^2$ diverges for $t\to-\infty$. }

From Lemma 2, if the space-time is incomplete $H/a$ diverges and therefore, from Lemma 5, $\dot{H}/a^2$ also diverges. Hence, space-times in which $H\to0$ are either complete or have a parallelly propagated curvature singularity.

Finally, let us take one example of physical interest which are the cyclic Universes \cite{IS19,IS21} with scale factor $a=P(t)\e ^{Nt/T}$, being $P(t)$ a periodic function with period $T$. They satisfy condition \textbf{(i)} and are incomplete. Notwithstanding, being $H(t)=N/T+\dot{P}/P$ a periodic function w.r.t. the cosmological time, its limit for $t\to-\infty$ does not exist, as it oscillates not approaching any definite value. This means that our extension does not work: if the Hubble function does not have a definite limit for $\xi\to\xi_{*}$, it cannot be continuously extended past this boundary. In fact, being the curvature scalars proportional to $H^2$, no $\mathcal{C}^{2}$ extension can be found through the past boundary, as there exists no continuous extension for the curvature invariants.

Physically, what happens is that the proper time $\mathcal{T}(t)$ elapsed during a period $T$ as a function of the cosmological time, which is given by

\begin{equation}
    \mathcal{T}(t)=\int^{t+T}_{t}\frac{\rmd t}{\sqrt{1+v_0^2/a^2}},
\end{equation}
goes to zero when $t\to-\infty$. To see this, take $a(t)^2\ll v^2_0$ and we have that

\begin{equation}
    \mathcal{T}(t)\approx\frac{1}{|v_0|}\int^{t+T}_{t}a\rmd t
\end{equation}
Being $P(t)$ periodic, then $ P_{\min}\leq P\leq  P_{\max}$, and

\begin{equation}
     \frac{T}{N} P_{\min} \e^{Nt/T}(\e^{N}-1) \leq\int^{t+T}_{t}a\rmd t\leq \frac{T}{N} P_{\max} \e^{Nt/T}(\e^{N}-1),
\end{equation}
so that

\begin{equation}
    \lim_{t\to-\infty}\mathcal{T}(t)= \frac{1}{|v_0|}\lim_{t\to-\infty}\int^{t+T}_{t}a\rmd t=0,
\end{equation}
by the bounding theorem. Therefore, the non-comoving observaber sees the Universe oscillating with a divergent frequency when $\xi\rightarrow\xi_{*}$. 

\section{Conclusion}
\label{sec_conclusion}
The results obtained in this paper can be summarized in the following theorem:

\textbf{Definition 3.} \textit{A spatially flat FLRW space-time satisfying assumptions \textbf{(i)}, \textbf{(ii)}, \textbf{(iii)} is said to be past-asymptotically de Sitter.} 

\vspace{0.4cm}
\textbf{Theorem 1.} \textit{Let a spatially flat FLRW space-time in which $a\to0$ in the past boundary. It has a $\mathcal{C}^2$ extension through such boundary if and only if it satisfies the following conditions:
\begin{enumerate}
    \item Cosmological time goes up to minus infinity, i.e., $t\in(-\infty,t_0]$.
    \item It is past-asymptotically dS , as in Definition 3.    
\end{enumerate}
    Moreover, the boundary is a local minimum of the squared scale factor and there is a bounce.}

\vspace{0.2cm}
\textit{Proof:} The space-time must necessarily fulfill assumption \textbf{(i)}, on the contrary, if $t\in [t_{in},t_0]$, with $t_{in}$ finite and $a(t_{in})=0$, then there is a scalar curvature singularity \cite{N5}.
If assumption \textbf{(ii)} is not fulfilled then, in the asymptotic limit, $H$ either diverges, implying a scalar curvature singularity, or it is zero, implying a parallelly propagated singularity as shown in Corollary 1 of Lemma 5, or the limit does not exist and no extension is possible, as exemplified by the cyclic model discussed above. Finally, assumption \textbf{(iii)} is self-evident, in order to avoid parallelly propagated singularities. Hence space-time must be past-asymptotically dS, and the extension can be made using the non-comoving coordinates given by (\ref{tangent_non-comoving_observer}) and (\ref{spatialtrans}), which is $\mathcal{C}^2$ if and only if $\dot{H}/a^2$ has a definite limit and $H\to\mbox{const}\neq0$ as $t \rightarrow -\infty$. Moreover, it must contain a bounce, as discussed.

Further, higher regularity in the extension can be obtained following the same reasoning in the Theorem 3.15 in \cite{N5}. Namely, in the past boundary the metric components and its derivatives up to second order are all products of $a$, $H$, $\dot{H}$, and $\dot{H}/a^2$. Thence, $\mathcal{C}^{k+2}$ regularity, with $k>0$, is assured whenever these functions are $\mathcal{C}^{k}$.

Notice that, when an extension is possible, the non co-moving observer experiences a \textit{kinematic bounce} exclusively due to its velocity, and hence there is no need for violation of the null energy condition: it moves in the locally contracting part reaching the $a=0$ boundary, emerging in the expanding region accessible to the co-moving observer after the bounce. This one only experiences the eternal expanding phase. Note that the spatial geometry is a curved inhomogeneous and anisotropic hyper-surface, and the bounce occurs in different proper times for each observer. Therefore, no violation of the null energy condition is necessary. Some examples are de Sitter space-time itself (obviously), and the toy model with $a\propto\mbox{sech}(\alpha t)$. Of course, there are also models which are extensible according to Theorem 1, but violate the null energy condition. For example, assuming the leading term in the energy density generating the space-time curvature to be a cosmological constant, sub-leading terms with constant equation of state must violate the null energy condition in order to have a finite $\dot{H}/a^2$ in the past-boundary, see the discussion in \cite{YQ18}.

Furthermore, while the cyclic Universe proposed in Refs. \cite{IS19,IS21} is indeed incomplete, the Penrose's Conformal Cyclic Cosmology (CCC) model \cite{P10} might be extended using our framework. The arguments used in \cite{N4} to sustain the incompleteness of the CCC model are based in a misunderstanding that non co-moving observers see a dS space-time with a homogeneous and isotropic spatial section with negative curvature, hence with a vanishing co-moving volume. However, as discussed in this paper, the non co-moving geodesics sees a very different spatial geometry, with a never vanishing co-moving volume, and can be completed.  

Note that a past asymptotically dS geometry, as in Definition 3, does not necessarily have all the symmetries of a dS space-time. Possible physical realizations are scalar fields with the non-singular potentials given in Ref. \cite{YQ18}. There, conditions are given for slow-roll inflationary models, that fulfill assumptions (i) and (ii), to satisfy assumption (iii). For example, it is shown that a Starobinsky model has a p.p. curvature singularity, while a Higgs-like potential can be extended. Anyway, whatever scenario one might have in mind, it seems to us quite premature to discard the extension described in this paper, and others, with arguments of implausible initial conditions. There is no sufficient knowledge of Quantum Gravity and Quantum Cosmology, which is expected to be the theory of initial conditions for Cosmology \cite{hartle}, to say much about this fundamental and contrived issue.

As generalizations of the present work, one first step should be to investigate its extension to Friedman models with curved space-like hypersurfaces. For the more involved anisotropic and non-homogeneous cases, the generalized Hubble factor $\mathcal{H}_{BGV}$, and the space-like volume of space-time using an appropriate choice of time may be useful tools to work with, but this will be the subject of future work.

\begin{acknowledgments}
\textit{Acknowledgments}-- ALFJ was funded by the Research Support Foundation of Esp\'{\i}rito Santo - FAPES grant number 13/2019 and by the Research Support Carlos Chagas Foundation of Rio de Janeiro - FAPERJ. NPN acknowledges the support of CNPq of Brazil under grant PQ-IB 310121/2021-3. VNX acknowledges the financial support of Coordenação de Aperfeiçoamento de Pessoal de Nível Superior – Brasil (CAPES) – Finance Code 001.
\end{acknowledgments}


\end{document}